\documentclass[pra,twocolumn,amsmath, superscriptaddress,notitlepage,showpacs]{revtex4-1}
\usepackage{amsmath,amsfonts,amssymb,amstext,amscd,amsthm,dsfont,bbm,natbib,braket}
\usepackage[T1]{fontenc}
\usepackage{physics}
\usepackage{color}
\usepackage{soul,xcolor}
\usepackage{gensymb}
\usepackage{subcaption}
\usepackage[colorlinks, breaklinks, urlcolor={cyan}, linkcolor={red}, citecolor={blue}]{hyperref}
\usepackage[normalem]{ulem}
\usepackage{graphicx}
\usepackage{bbold}
\usepackage{braket}
\usepackage{placeins}
\usepackage{float}

\usepackage{ulem}

\begin{document}
\title{Realizing the Petz Recovery Map on an NMR Quantum Processor}

\author{Gayatri Singh}
\affiliation{Department of Physical Sciences, Indian Institute of Science Education \& Research Mohali,
Sector 81 SAS Nagar, Manauli PO 140306 Punjab India}

\author{Ram Sagar Sahani}
\affiliation{Department of Physical Sciences, Indian Institute of Science Education \& Research Mohali,
Sector 81 SAS Nagar, Manauli PO 140306 Punjab India}

\author{Vinayak Jagadish}
\affiliation{Department of Computer Science and Engineering, Amrita School of Computing, Amrita Vishwa Vidyapeetham, Amritapuri 690525, India}
\affiliation{National Institute for Theoretical and Computational Sciences (NITheCS), Stellenbosch, 7600, South Africa}
\affiliation{Centre for Quantum Science and Technology, Chennai Institute of Technology, Chennai, 600069, India} 

\author{\ Lea Lautenbacher}
\affiliation{Institut f\"ur Theoretische Physik, Albert-Einstein-Allee 11, Universit\"at Ulm, D-89069 Ulm, Germany}

\author{Nadja K. Bernardes}
\affiliation{Departamento de F\'{\i}sica, Universidade Federal de Pernambuco, Recife, PE  50670-901 Brasil}

\author{Kavita Dorai}
\affiliation{Department of Physical Sciences, Indian Institute of Science Education \& Research Mohali,
Sector 81 SAS Nagar, Manauli PO 140306 Punjab India}
\begin{abstract} 
The Petz recovery map is a central construct in quantum information theory, providing an explicit, channel-aware prescription for reversing the effects of noise. Unlike standard quantum operations, the Petz map is intrinsically dependent on a chosen reference state, which makes its physical implementation and experimental validation particularly challenging. Here, we report an experimental realization of Petz recovery maps on a nuclear magnetic resonance (NMR) quantum processor using the duality quantum computing (DQC) algorithm. We investigate two paradigmatic single-qubit noise models: amplitude damping and phase damping, and construct corresponding families of Petz recovery maps for varying reference states. By systematically tuning the reference state, we experimentally demonstrate the state-adapted nature of Petz recovery, observing both enhanced recovery when the reference state is well matched and fidelity degradation for mismatched choices. Our experimental results show close quantitative agreement with theoretical predictions, providing direct evidence that the Petz recovery map constitutes a physically realizable, reference-state-dependent recovery channel rather than a purely formal inverse of noise. This work bridges the gap between the abstract information-theoretic formulation of Petz recovery and its implementation on a realistic quantum platform, and establishes an experimental benchmark for testing noise-adapted recovery strategies on near-term quantum devices.
\end{abstract}
\maketitle  
\section{Introduction}
Open quantum systems~\cite{Davies1976-ab,Haroche2006-jo} are defined by their
interaction with an external environment, which breaks the idealization of a
perfectly isolated/closed quantum system. While classical systems can exchange
energy with their surroundings in a straightforward manner, quantum systems
exhibit a far more delicate sensitivity to environmental interactions. This
sensitivity manifests in phenomena such as decoherence, wherein a quantum
system initially in a pure state evolves into a statistical mixture due to the
build-up of entanglement with the environment. Decoherence is particularly
detrimental to quantum technologies, as it erodes the fundamental quantum
properties, namely superposition and entanglement, that underpin quantum
computing, communication, sensing
etc.~\cite{Chuang1995,Shor1995,Aschauer2002,Matsuzaki2011,Albash2015,Schlosshauer-Selbach2007}

Mathematically, the process of decoherence can be understood through the lens of quantum channels, which describe the dynamics of the quantum system of interest under noise. For instance, a dephasing channel suppresses off-diagonal elements of the density matrix, effectively randomizing quantum phases, while an amplitude damping channel models energy dissipation, such as spontaneous emission in atomic systems. These noise processes are not merely theoretical constructs; they are omnipresent in experimental platforms, from superconducting qubits to trapped ions and NMR systems~\cite{Vandersypen2005,Singh2020,Deslauriers2006}.

To counteract decoherence and information loss in quantum systems, a variety of error mitigation and recovery strategies have been developed. These broadly include control-based approaches that utilize measurements to apply real-time corrections \cite{ zhang-ieee-2010, cao-pra-2017}, as well as schemes involving probabilistic elements \cite{Wang-pra-2014, Harraz-2020, Kim2012, gayatri-arxiv-2024}. While these methods can effectively suppress decoherence, and achieves higher fidelities at the cost of reduced success probabilities.
 Another widely studied framework for protecting quantum information is quantum error correction (QEC) \cite{ ryan-prx-2021,Gicev-prr}.  In standard QEC protocols, logical qubits are encoded into multiple physical qubits, and errors can be detected through syndrome measurements. These are subsequently corrected using appropriate operations. It typically requires significant resource overhead in terms of additional qubits and repeated syndrome measurements. This motivates the exploration of alternative,
channel-adapted recovery strategies that can operate directly at the physical level.

The Petz recovery map \cite{Petz1986-by, Petz2003-vy} provides such a strategy.  It is a channel-adapted recovery operation, which arises naturally from the saturation of the data-processing inequality for quantum relative entropy and therefore represents an information-theoretically motivated recovery map. For a fixed noise channel and reference state, it serves as a canonical recovery operation and is optimal in an information-theoretic sense, with achievable recovery fidelity characterizing the best possible performance attainable under that noise model~\cite{bai-prl-2025}. 
It is worth emphasizing that the Petz recovery map
differs conceptually from standard diagnostic tools such
as quantum process tomography and randomized bench-
marking. While these methods aim to characterize noise,
the Petz map instead directly addresses the recoverability of quantum information under a given channel. In
this sense, it provides a complementary, operational perspective: rather than identifying the noise, it quantifies how well its effects can be reversed. 

\color{black} 
 
Despite its theoretical importance, and applications in areas such as QEC, black hole physics,  quantum resource theories, and quantum thermodynamics~\cite{Barnum2002-lv,Ng2010-eh,Aw2021-pn,Kwon2022-mo,Surace2023-wj}, experimental realizations of the Petz recovery map remain limited.  Its implementation is
challenging due to its explicit dependence on the noise
channel, its adjoint, and a tunable reference state. Only recently, an implementation of the Petz map in trapped-ion platforms has been achieved~\cite{png-pra-2025}, marking a step toward its practical realization. In parallel, algorithmic approaches have been developed to embed the Petz recovery channel within quantum circuits~\cite{Gilyen2022-li}, and noise-adapted recovery strategies inspired by the Petz map have been designed to achieve near-optimal fidelity for arbitrary quantum codes and noise models~\cite{Biswas2024-tk}.

The central goal of this work is to bridge the gap between the theoretical foundations of the Petz recovery map and its practical implementation in realistic quantum settings. We propose an experimental realization of Petz maps using NMR techniques. Building on the detailed theoretical analyses of the Petz map under physically relevant noise models~\cite{Lautenbacher2022-qj,Lautenbacher2024-yu}, we focus on two paradigmatic channels, dephasing and amplitude damping, which are both theoretically well understood and naturally realized in NMR platforms.  In 
general setting, the experimental implementation necessitates resource-intensive subroutines such as quantum singular value transformation and block encoding \cite{Biswas2024-tk}. To circumvent these complexities, we employ the ancilla-assisted DQC algorithm~\cite{xin-pra-2017} for the NMR simulation of the dynamics. While this framework has previously been used to implement noise channels themselves, here it serves a fundamentally different purpose: enabling the realization of recovery maps. The ancilla-assisted implementation is essential, as non-unitary noise processes cannot, in general, be inverted by system-only unitaries.

\color{black}
Our experimental results show excellent agreement with theoretical predictions, providing a compelling validation for the physical implementability of Petz recovery map. Beyond this validation, our work establishes a practical experimental framework for the deployment of reference-dependent Petz recovery map in real-world quantum protocols. By explicitly revealing both its capabilities and limitations, this work bridges the gap between the abstract, information-theoretic formulation of Petz recovery and its practical realization in a laboratory setting. We anticipate that these results will inform future investigations of noise-adapted recovery strategies and provide a benchmark for testing information-theoretic recovery schemes on near-term quantum platforms.

This paper is structured as follows. We begin by introducing the theoretical
background of quantum channels and the Petz map. We then outline the quantum channels and the experimental setup used
in our NMR implementation. Our main results are followed by a detailed
comparison with the theoretical predictions. Finally, we conclude with a
discussion of the implications of our findings and potential future directions.

\section{Quantum Channels and Petz recovery maps}
The evolution of an open quantum system is described by a \textit{quantum channel}. A quantum channel~\cite{Jagadish2018-ya} is typically represented by a \textit{completely positive, trace-preserving (CPTP)} map. Complete positivity (CP) ensures that the map maintains the positivity of quantum states when acting on a subsystem of a larger extended system, thereby preserving the physical validity of the overall state. Trace Preservation (TP)  ensures that the total probability is conserved during the dynamics.

Mathematically, a quantum channel \(\Lambda\) is a linear map acting on the space of density matrices \(\mathcal{B}(\mathcal{H})\), the space of bounded linear operators on a Hilbert space \(\mathcal{H}\). The quantum state \(\rho\) of the system of interest transforms as
\begin{equation}
    \rho' = \Lambda(\rho),
\end{equation}
where \(\rho'\) is the new state after the action of the channel \(\Lambda\) on \(\rho\).
Quantum channels admit the \textit{operator sum representation}, also known as the \textit{Kraus representation}. In this representation, a quantum channel \(\Lambda\) is described by a set of operators \(\{\mathcal{K}_m\}\), known as the \textit{Kraus operators},
\begin{equation}
\label{eq:Kraus}
    \Lambda(\rho)=\sum_{m=0}^{d^2-1} \mathcal{K}_m\rho \mathcal{K}_m^\dagger,
\end{equation}where $d$ is the dimension of the Hilbert space of the system. It must be noted that the Kraus representation is not unique, and is always defined up to a unitary. In the canonical representation, one could have a maximum of $d^2$ Kraus operators.
The map is trace preserving if the set of operators $\{\mathcal{K}_m\}$ satisfy 
\begin{equation}\label{eq:trace}
    \sum_m \mathcal{K}_m^\dagger \mathcal{K}_m = \mathbb{1},
\end{equation}
where \(\mathcal{K}_m^\dagger\) is the Hermitian conjugate of \(\mathcal{K}_m\), and \(\mathbb{1}\) is the identity operator on the Hilbert space. 

Given two Hilbert spaces $\mathcal{H}_A$ and $\mathcal{H}_B$, we define $S(\mathcal{H}_A)$ and $S(\mathcal{H}_B)$ as sets of density operators that act on systems $A$ and $B$, respectively. A quantum channel $\Lambda \colon S(\mathcal{H}_A) \to S(\mathcal{H}_B)$ is said to be \emph{reversible} if there exists a quantum channel $\mathcal{R} \colon S(\mathcal{H}_B) \to S(\mathcal{H}_A)$, called the \emph{recovery channel}, such that for all density operators $\rho \in S(\mathcal{H}_A)$, the following holds:
\begin{equation}
(\mathcal{R} \circ \Lambda)(\rho) = \rho.
\end{equation}

To quantify the extent to which a quantum channel affects the distinguishability of input states, we often turn to the concept of relative entropy. Under the action of a quantum channel $\Lambda$, the relative entropy between two states can never increase. This result is known as the \emph{monotonicity of relative entropy}, or the \emph{data processing inequality}, and is expressed as:
\begin{equation}
D(\Lambda(\rho) \|\Lambda(\sigma)) \leq D(\rho \|\sigma),
\end{equation}
where $D(\rho \|\sigma) = \text{Tr}(\rho \log \rho) - \text{Tr}(\rho \log \sigma)$ denotes the relative entropy between the states $\rho$ and $\sigma$.

If there exists a recovery channel $\mathcal{R}$ such that $(\mathcal{R} \circ \Lambda)(\rho) = \rho$ and $(\mathcal{R} \circ \Lambda)(\sigma) = \sigma$, then applying monotonicity again to the channel $\mathcal{R}$ yields the condition of equality,
\begin{equation}
D(\rho \|\sigma) = D(\Lambda(\rho)\|\Lambda(\sigma)).
\end{equation}
This equality characterizes the situation where perfect recovery is possible. In fact, Petz~\cite{Petz1986-by,Petz2003-vy} identified an explicit recovery map $\mathcal{R} = \mathcal{P}_{\sigma, \Lambda}$, given by
\begin{equation}
\label{petz}
\mathcal{P}_{\sigma, \Lambda}(\cdot) = \sigma^{1/2} \Lambda^\dagger \left[ {\Lambda(\sigma)}^{-1/2} (\cdot) {\Lambda(\sigma)}^{-1/2} \right] \sigma^{1/2},
\end{equation}
where $\Lambda^\dagger$ denotes the adjoint of the channel $\Lambda$ and $\sigma$ is the \textit{reference state}. 

The reference state $\sigma$ encodes prior information about the class of states for which recovery is expected to be effective. Consequently, the Petz map is not a universal inverse of the channel $\Lambda$, but rather a state-adapted recovery map that achieves exact or near-optimal recovery
for those states that (approximately) saturate the data processing
inequality with respect to the chosen reference state. This state-adapted nature of the Petz map has been emphasized in recent works~\cite{Lautenbacher2022-qj,Lautenbacher2024-yu} and is central to its operational interpretation.

In this work, we study the performance of the Petz map when applied to the following two quantum channels.

\paragraph{Amplitude Damping (AD) Channel:}
An AD channel models energy dissipation from the quantum system of interest to the environment and drives the system towards the ground state $\ket{0}$, affecting both populations and coherences of the density matrix. The action of the AD channel on a qubit can be described  by the following Kraus operators:
\begin{equation}
        \mathcal{K}_0^{AD}=\begin{pmatrix}
    1&0\\
    0& \sqrt{1-p}
\end{pmatrix} \hspace{0.4cm};\hspace{0.4cm}
\mathcal{K}_1^{AD}=\begin{pmatrix}
    0&\sqrt{p}\\
    0& 0
\end{pmatrix},\label{ad-kraus}
\end{equation}
where $p\in[0,1]$ denotes the channel strength. 

Motivated by the asymmetric nature of the AD channel, where the $\ket{0}$ state
remains unaffected while the $\ket{1}$ state undergoes a probabilistic decay,
we consider a parameterized reference state that is
diagonal in form. With this specific choice of reference state,
$\sigma=(1-\epsilon)\ket{0}\bra{0}+\epsilon\ket{1}\bra{1}$, the corresponding
Kraus operators for the Petz recovery map for AD channel are evaluated to be
\begin{equation}
    \begin{split}
    &\mathcal{M}_0^{AD}=\begin{pmatrix}
        \sqrt{\frac{1-\epsilon}{1-(1-p)\epsilon}}&0\\
        0&1
    \end{pmatrix}\\
     &\mathcal{M}_1^{AD}=\begin{pmatrix}
     0& 0\\
     \sqrt{\frac{{p\epsilon}}{1-(1-p)\epsilon}}&0
    \end{pmatrix}.
    \end{split}\label{adpetz}
\end{equation}

\paragraph{Phase Damping (PD) Channel:}
A PD channel causes the loss of quantum coherence by suppressing the off-diagonal elements of the density matrix of the qubit, thus erasing relative phase information between the computational basis states, without altering their populations (diagonal elements). In its Kraus representation, the PD channel on a qubit is described by 
\begin{equation}\label{pd-kraus}
        \mathcal{K}_0^{PD}=\sqrt{1-\frac{p}{2}}\begin{pmatrix}
    1&0\\
    0& 1
\end{pmatrix} \hspace{0.3cm};\hspace{0.3cm}
\mathcal{K}_1^{PD}=\sqrt{\frac{p}{2}}\begin{pmatrix}
    1&0\\
    0& -1
\end{pmatrix},
\end{equation}
where $p\in[0,1]$ is the strength of the channel. Note that for $p=1$ the coherences of the density matrix vanish completely. The PD channel has the most significant effect on superposition states, while it leaves diagonal states unaffected. Therefore, to construct the Petz recovery map for this channel, we consider a reference state of the form $\sigma=(1-\epsilon)\ket{+}\bra{+}+\epsilon\ket{-}\bra{-}$, where $\ket{\pm}=\ket{0}\pm\ket{1}/\sqrt{2}$ are the eigenstates of the Pauli-$x$ operator.  Using this reference state, the corresponding Kraus operators for the Petz recovery map of the PD channel are given by
\begin{equation}\label{pdpetz-kraus}
\begin{split}
    \mathcal{M}_0^{PD}&=\begin{pmatrix}
    \lambda_+& \lambda_-\\
     \lambda_- &  \lambda_+
\end{pmatrix} \hspace{0.3cm} \text{with}\hspace{0.1cm}\lambda_{\pm}=\sqrt{1-\frac{p}{2}}\left(a\sqrt{1-\epsilon}\pm b\sqrt{\epsilon}\right)\\
\mathcal{M}_1^{PD}&=\begin{pmatrix}
    \mu_+&\mu_-\\
    -\mu_-& -\mu_+
\end{pmatrix}\hspace{0.3cm} \text{with}\hspace{0.1cm}\mu_{\pm}=\sqrt{\frac{p}{2}}\left(a\sqrt{\epsilon}\pm b\sqrt{1-\epsilon}\right),
\end{split}
\end{equation}
where $a$ and $b$ are evaluated to be
\begin{equation*}
\begin{split}
        a=&~\frac{1}{\sqrt{2} \sqrt{p \epsilon+(2-p) (1-\epsilon)}}\\
        b=&~\frac{1}{\sqrt{2} \sqrt{p -p\epsilon+\epsilon(2-p)}}.
\end{split}
\end{equation*}

We emphasize that the reference states chosen for the  AD
and PD channels are not arbitrary nor are they intended to yield uniform recovery for all possible input states. Instead, they are physically motivated by the structure of the respective noise channels. This specific parametrization serves as a methodological choice to illustrate the central feature of the Petz recovery map. In the case of amplitude damping, the channel possesses a
preferred energy eigenbasis and a fixed point corresponding to the ground state, motivating a reference state diagonal in the
computational basis. For phase damping, which selectively suppresses coherences while preserving populations, a reference state diagonal in the Pauli-$X$ basis captures the coherence structure most strongly affected by the channel.  As a result, the corresponding Petz recovery
maps are expected to perform optimally only for input states that have
significant overlap with the chosen reference state.


\section{Experimental Details} 
\label{experiment}
\subsection{Realization of NMR 
qubits}

\begin{figure}
\centering
\includegraphics[width=1\linewidth]{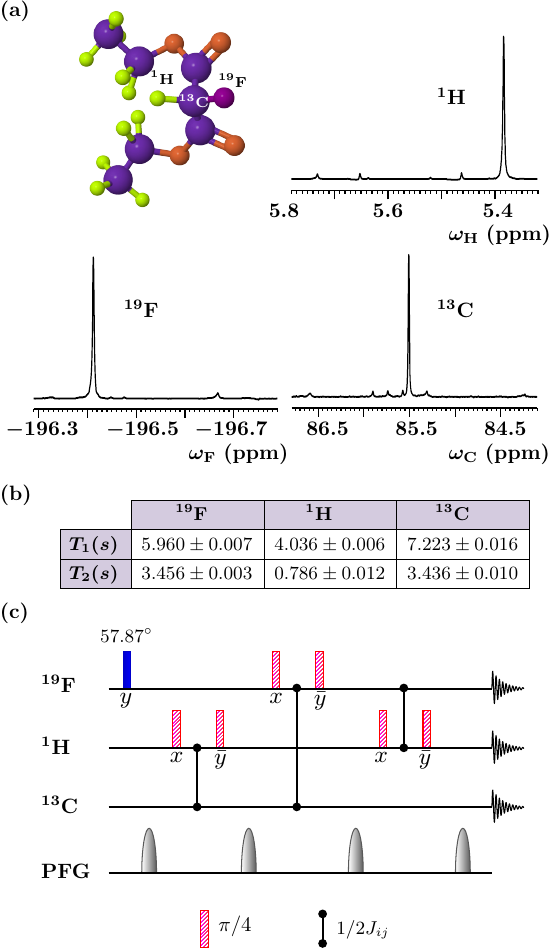}
\caption{(Color online) $\rm(a)$ The molecular structure of
$^{13}\rm C$ labeled diethyl
fluorormalonate, used as a three-qubit system, with 
NMR-active nuclei $^1$H, $^{19}$F, and $^{13}$C. The NMR spectra of the three nuclei acquired after applying a $\pi/2$ readout
pulse to the PPS $\ket{000}$.  The
horizontal axis indicates frequency in parts per million (ppm).
$\rm(b)$ Longitudinal relaxation ($T_1$) and transverse relaxation
($T_2$) times (in seconds) of different nuclei.
$\rm(c)$ NMR pulse sequence for preparing the three-qubit PPS. Rectangular blocks represent
RF pulses, with the flip angle and phases labeled beneath each pulse. An overbar indicates
negative phase.  The solid blue rectangle denotes a rotation along the $y-$axis by
an angle $57.87\degree$. Cross-hatched magenta rectangles denote $\pi/4$
pulses, while  gray-shaded shapes indicate pulsed field gradients
(PFG). A vertical line connecting two solid dots between qubits $i$ and
$j$ indicates free evolution under the NMR Hamiltonian for a duration
of $1/2J_{ij}$, where $J_{ij}$ is the scalar J-coupling.
}
\label{fig1-hfc-pps}
\end{figure}

We used a three-qubit NMR quantum processor to
experimentally realize quantum channels and their associated Petz recovery
maps.  The three spin-half nuclei namely, $^1$H, $^{19}$F, and $^{13}$C, in a
$^{13}$C-labeled diethyl fluoromalonate molecule dissolved in acetone D6, were
assigned as three individual qubits (Fig.~\ref{fig1-hfc-pps}$\rm(a)$).   All
the experiments were carried out at room temperature
($\sim$300 K) on a Bruker Avance-III 600 MHz NMR
spectrometer equipped with a 5 mm quadruple resonance inverse (QXI) probe. The durations of $\pi/2$ pulses for $^1$H, $^{19}$F, and $^{13}$C
were $9.3\,\mu$s $, 23.5\,\mu$s and $15\,\mu$s at power levels of $18.14$ W,
$42.47$ W, and $179.47$ W, respectively.

The internal Hamiltonian of the three-qubit
system in the rotating frame, under the weak coupling approximation, 
is expressed as
\begin{equation}
\mathcal{H}=-\sum_{i=1}^3 (\omega_{i}-\omega_i^{rf}) I_{i z}+\sum_{i>j,i=1}^3 J_{i j} I_{i z} I_{j z}.\label{eq_ham},
\end{equation}
where the indices $i,j$ label the qubits.  In
our case, the nuclei $^{19}$F, $^{1}$H and $^{13}$C are qubits 1,2, and 3
respectively; $\omega_i$ denotes the chemical shift, $\omega_i^{rf}$ is the
rotating frame frequency, $I_{iz}$ is the $z$-component of the spin
angular momentum of the $i$th nucleus and $J_{ij}$ is the
scalar coupling between the $i$th and $j$th nuclei, with
experimentally measured values : $J_{\rm CH}=161.42$ Hz, $J_{\rm
FH}=47.50$ Hz, and  $J_{\rm FC}=-191.90$ Hz. In NMR, the recovery of
longitudinal magnetization is characterized by the $T_1$ relaxation time, while the decay of transverse magnetization is governed by
the $T_2$ relaxation time. The experimentally measured values of $T_1$
and $T_2$ relaxation times in seconds for the different
nuclei given in
Fig.~\ref{fig1-hfc-pps}($\rm b$).

We initialize our system in the pseudopure state (PPS) starting from
thermal equilibrium via the spatial averaging
technique~\cite{oliveira-ch4},   which comprises single qubit
rotations, free evolution and pulsed field gradients (PFG).
A PPS retains the spectral properties of a pure state.
The NMR pulse sequence to prepare the PPS is depicted in
Fig.~\ref{fig1-hfc-pps}$\rm(c)$  and the corresponding density matrix is
expressed as
\begin{equation}
\rho_{_{\text{PPS}}}=\frac{1-\kappa}{2^3}
\mathbb{1}^{\otimes 3}+ 
\kappa \ket{000}\bra{000},
\end{equation}
where $\kappa$ is the thermal polarization factor
(typically  $\sim 10^{-5}$) 
at room temperature. The term $\mathbb{1}$ denotes the $2\times 2$
identity operator. The PPS was experimentally reconstructed using the the least
squares constrained convex optimization method \cite{gaikwad-qip-2021} with a
set of tomographic pulses
$\{\mathbb{1}\mathbb{1}\mathbb{1},\mathbb{1}\mathbb{1}R_y,\mathbb{1}R_yR_y,R_y\mathbb{1}\mathbb{1},R_xR_yR_x,R_xR_xR_y,R_xR_xR_x\}$.
Here $R_{x(y)}$ denotes a spin-selective $\pi/2$ rotation
about the $x(y)$ axis. The fidelity of the
experimentally reconstructed state $\mathrm {\rho_{exp}}$ with respect to the
expected theoretical state $\mathrm {\rho_{th}}$ was evaluated using the
Uhlmann-Jozsa fidelity measure~\cite{uhlmann-rpmp-1976,jozsa-jmo-1994} as
\begin{equation}
\mathcal{F}(\rho_{\mathrm{exp}},
\rho_{\mathrm{th}})= \left[\text{Tr}\left(\sqrt{\sqrt{\rho_{\mathrm{th}}}.\rho_{\mathrm{exp}}.\sqrt{\rho_{\mathrm{th}}}}\right)\right]^2.
\end{equation}
The average state fidelity of 
the experimentally reconstructed PPS was computed
to be $0.9799\pm0.0010$.

\subsection{Experimental Simulation using the Duality Quantum Computing (DQC)
Algorithm}
\label{sec-dqa}

\begin{figure*}[htb!]
\centering
\includegraphics[scale=1]{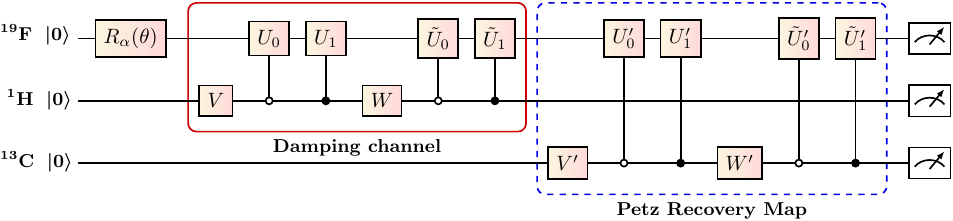}
\caption{(Color online) Schematic of the quantum circuit
used to implement the damping channel  and the corresponding Petz
recovery map.  The gate
$R_\alpha(\theta)$ denotes a rotation by an
angle $\theta$ about the $\alpha\in\{x,y,z\}$ axis, used to
initialize the system qubit ($\rm^{19}F$) in state $\ket{\Phi}$. The
red solid rectangle denotes the segment of the circuit implementing
the damping channel, comprising unitary gates $V,W$ and controlled
unitary operations $U_0,U_1,\tilde{U}_0$, and $
\tilde{U}_1$. The Petz
recovery map, shown within the blue dashed region, involves the
unitaries $V',W'$ and controlled unitaries $U_0',U_1',\tilde{U}'_0$,
and $ \tilde{U}'_1$.   The specific forms of these unitaries differ depending
on the channel (PD or AD) and are provided in Eqs.
\eqref{ad-dqa-unit1},\eqref{ad-dqa-unit2}, \eqref{pd-dqa-unit1}, and \eqref{pd-dqa-unit2}, respectively.
}
\label{fig2-quantcirc}
\end{figure*}

Each Kraus operator $\mathcal{K}_m$ corresponding to the quantum channel $\Lambda$ [as in Eq.~\eqref{eq:Kraus}]
can be decomposed as
\begin{equation}
\mathcal{K}_m = \tilde{U}_m \mathcal{L}_m,
\end{equation}
where, $\tilde{U}_m$ is a unitary operator and $\mathcal{L}_m$ is a
generally non-unitary operator \cite{wei-2018}. The index $m$
labels the individual Kraus operators and is not summed over. This decomposition can be understood as a polar-type
decomposition of the Kraus operator, in which the non-unitary action is
separated from a unitary correction.
The operator $\mathcal{L}_m$ can then be expressed as a linear combination of at most $d^2$ unitary operators $U_j$ acting on the $d$-dimensional system Hilbert space,
\begin{equation}
\mathcal{L}_m=\sum_{j=0}^{d^2-1} \beta_j^m U_j,
\end{equation}
with complex coefficients $\beta_j$. 

The introduction of the unitary $\tilde{U}_m$ plays a
crucial role in the duality quantum computing (DQC) framework. The DQC algorithm enables the coherent implementation of linear combinations of unitary operators, but does not directly implement arbitrary non-unitary Kraus operators. By first realizing the action of
$\mathcal{L}_m$ through a linear combination of unitaries, and then applying the unitary correction $\tilde{U}_m$, the full Kraus operator $\mathcal{K}_m$ is implemented exactly.
As a result, each Kraus operator is simulated within a distinct subspace of the ancillary system  of at most $n=\log_2 d^2$ qubits~\cite{xin-pra-2017,gulati2024,gayatri-arxiv-2024}.

The steps in the DQC algorithm are delineated below.

\begin{itemize} 
\item The system qubit is initialized in the desired input state $\ket{\Phi}$, while the ancillary system consisting of $n$ qubits is initialized in the $\ket{0}^{\otimes n}$
state.  
\item  Following this, a unitary operator $V$
is implemented on the ancillary qubit and the
joint state of the system and ancillary qubit,
evolving the initial state as
$\ket{\Phi}\ket{0}\rightarrow\sum_{j=0}^{d^2-1}
\ket{\Phi}V_{j0}\ket{j}$.  Here, $V_{j0}$
represents the first column of the unitary
matrix $V$, determined by the complex
coefficients in the unitary expansion of the Kraus
operator. The remaining columns of $V$ are
constructed using the Gram–Schmidt
orthogonalization procedure to ensure
unitarity.  
\item A controlled unitary
operation $\sum_{j=0}^{d^2-1}U_j
\otimes \ket{j} \bra{j}$ is performed
on the system qubit with the state of
the ancillary qubit acting as control,
which transforms the state as 

\begin{equation}
\sum_{j=0}^{d^2-1} V_{j0} \ket{\Phi}
\ket{j} \rightarrow \sum_{j=0}^{d^2-1} V_{j0} U_j \ket{\Phi} \ket{j}. 
\end{equation} 
\item Next, a unitary operation $W$ is applied on the ancillary qubit, which leads to the state evolution
\begin{equation} \label{duality-gate}
   \sum_{m,j=0}^{d^2-1}
 W_{mj}V_{j0} U_j\ket{\Phi} \ket{m}= \sum_{m=0}^{d^2-1} \mathcal{L}_m\ket{\Phi}\ket{m}.
\end{equation}
The elements of the matrix $W$ are uniquely determined 
 based on the choice of the unitary matrix $V$ and 
are constructed such that
the operator $\mathcal{L}_m$ satisfies 
the relation
$\mathcal{L}_{m}=\sum_{j=0}^{d^2-1} W_{m j} V_{j0} U_{j}$.
\item An additional controlled unitary operation is implemented on the system
qubit, with the state of the ancillary qubit as control, to map the operator
$\mathcal{L}_m$ to the Kraus operators $\mathcal{K}_m$. This
transformation is expressed as 
\begin{equation}\label{kraus-sim}
\sum_{m=0}^{d^2-1} \tilde{U}_m\mathcal{L}_m 
\ket{\Phi}\ket{m}=\sum_{m=0}^{d^2-1} \mathcal{K}_m \ket{\Phi}\ket{m},
\end{equation}
where, each Kraus operator is defined as
$\mathcal{K}_m=\tilde{U}_m\sum_{j=0}^{d^2-1} W_{m j} V_{j0}
U_{j}$. In this step, the controlled application of $\tilde{U}_m$
ensures that the effective action on the system qubit exactly matches
the target Kraus operator $\mathcal{K}_m$.

\item In the final step, a measurement is performed on the ancillary system in
the computational basis. Upon measuring the ancillary qubit in
$\ket{m}$ state, we get the action of the Kraus operator
$\mathcal{K}_m$ simulated on the system qubit. 
\end{itemize}

We now move towards
the experimental demonstration of the recovery of a quantum state that has
undergone decoherence due to a given damping channel—either PD or AD channel,
using the Petz recovery map. In NMR systems, amplitude damping and phase
damping correspond to intrinsic decoherence mechanisms characterized by the
relaxation times $T_1$ and $T_2$, respectively. In our simulations, these
effects are modeled as ideal quantum channels applied explicitly to the system,
rather than arising from the underlying physical noise naturally present in the
NMR environment. This approach is also essential because the experimental
implementation of the associated recovery maps 
requires explicit knowledge of
the map itself. It must be noted that for the specific quantum channels
considered in our study, a single ancillary qubit is sufficient to implement the
desired channel dynamics, as the maps are of rank-2 (two canonical Kraus
operators). By 
the same analogy, the simulation of the Petz recovery map
associated with the corresponding damping channel also requires only one
ancillary qubit.

In all 
the experiments, we consider $^{19}\rm F$ as the system qubit and
$^{1}\rm H$ and $^{13}\rm C$ as ancillary qubits to implement the damping
channel and the corresponding Petz recovery map, respectively. The experimental protocol, beginning with the PPS is detailed below.
\begin{itemize}
\item
The system and ancillary qubits are
initialized in the state $\ket{\Phi}\otimes\ket{00}$ by applying the rotation
gate $R_\alpha(\theta)$ on the system qubit;
$R_\alpha(\theta)$ denotes a
rotation about the axis $\alpha \in\{x,y,z\}$ 
by an angle $\theta$.  
\item The damping
channel is then implemented, followed by the corresponding Petz recovery map  
using
the appropriate NMR pulse sequences. 
\item Finally, quantum state tomography is
performed to reconstruct the output density matrix of the system qubit. 
\end{itemize}
In NMR,
seven tomography pulses are required to reconstruct the complete density matrix
of a three-qubit system \cite{leskowitz-pra-2004}. However, in our
implementation, we are specifically interested in the final state of the system
qubit alone. As described earlier, the structure of the DQC algorithm ensures
that each Kraus operator acts within a distinct subspace of the ancillary
qubit. Taking advantage of this structure, we reconstruct the state of the
system qubit by performing tomography using only a reduced set of pulses
$\{\mathbb{1}\mathbb{1}\mathbb{1},R_y\mathbb{1}\mathbb{1}\}$. 
The signal
integrals obtained from these measurements are summed to yield the density
matrix of the system qubit, which is mathematically equivalent to tracing out
the ancillary qubits from the three-qubit density matrix.

To better contextualize the experimental overhead, we compare the resources required for the Petz map with those of standard characterization techniques: QPT and RB. For a system of dimension $d$, QPT requires the preparation of $d^2$ input states and measurement of $d^2$ observables, leading to an overall scaling of order $O(d^4)$. While, RB relies on averaging over many random circuit realizations. In contrast, once the Kraus operators describing the noise channel are known or engineered, the Petz recovery map can be implemented
within the same DQC framework used to simulate the channel. The experimental overhead is therefore primarily determined by the number of Kraus operators $m$, requiring an ancillary register
of size $\lceil \log_2 m \rceil$ (at most $ \log_2 d^2$).

\color{black}

Although the joint evolution of the system qubit and the ancillary qubits is unitary, the effective dynamics of the system alone is non-unitary because the ancillae are used to mediate controlled information loss and are subsequently traced out. In this way, the ancilla qubits function as an engineered environment that enables the realization of quantum channels in a fully programmable manner. Once information has leaked from the system into these ancillary degrees of freedom, the resulting non-unitary channel cannot, in general, be inverted by any unitary acting on the system alone. Ancilla-assisted recovery is therefore essential for reversing open-system dynamics. Within this setting, the Petz recovery map provides a principled, channel-dependent recovery operation that explicitly exploits the structure of the noise channel, in contrast to heuristic or standard unitary reversal schemes, thereby motivating its experimental implementation.

The quantum circuit shown in Fig.~\ref{fig2-quantcirc}, provides a general
framework applicable to both the channels and their respective Petz recovery
maps using the duality quantum algorithm. Within this framework, the simulation
of each channel involves a specific set of unitary operators tailored to the
Kraus decomposition of a quantum channel. We now proceed towards the
unitary constructions and their corresponding pulse sequences for 
each channel.

\subsection{Amplitude Damping (AD) Channel 
and the Associated Petz Recovery Map } 
\label{sec-adexp}

To simulate the action of the AD channel using the duality quantum algorithm, we first express each Kraus operator in [Eq.~\eqref{ad-kraus}] as a linear combination of Pauli matrices as
\begin{equation}
\begin{split}
        \mathcal{K}_0^{AD}&=\frac{1+\sqrt{1-p}}{2}\mathbb{1}+\frac{1-\sqrt{1-p}}{2}Z\\
          \mathcal{K}_1^{AD}&=\frac{\sqrt{p}}{2}X (\mathbb{1}-Z).
\end{split}\label{ad-pauli}
\end{equation}
Based on this decomposition, we identify the unitary matrices $V$, $W$, $U_j$  and $\tilde{U}_j$ as
\begin{equation}
\begin{split}
     U_0&=\mathbb{1},\hspace{0.45cm} U_1=Z~,\hspace{0.45cm} \tilde{U}_0=\mathbb{1}, \hspace{0.45cm} \tilde{U}_1=X,\\
     V&=W=\begin{pmatrix}
       \sqrt{\frac{1+\sqrt{1-p}}{2}}&\sqrt{\frac{1-\sqrt{1-p}}{2}} \\
       \sqrt{\frac{1-\sqrt{1-p}}{2}}&-\sqrt{\frac{1+\sqrt{1-p}}{2}}
    \end{pmatrix}          
\end{split}\label{ad-dqa-unit1}
\end{equation}  
where, $\mathbb{1}$ is the $2\times 2$ Identity operator and $X,Z$ represents Pauli-$x,z$ matrices respectively. Following Eq.~\eqref{kraus-sim}, we infer that the duality algorithm encodes
the action of each Kraus operator within orthogonal subspaces of the ancillary
qubit. For instance, the action of the Kraus operator $\mathcal{K}_0^{AD}$ is
simulated within the $\ket{0}$ sub-space of ancillary qubit, while
$\mathcal{K}_1^{AD}$ is simulated in $\ket{1}$ subspace. 

\begin{figure}[htb!]
\centering
\includegraphics[width=1\linewidth]{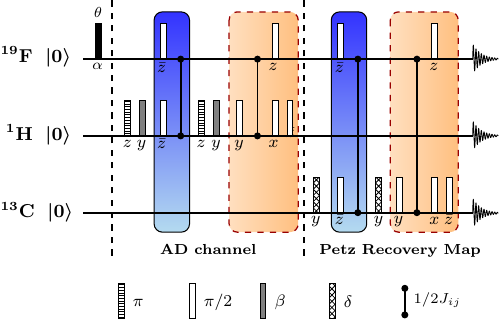}
\caption{(Color online) 
NMR pulse sequence used to realize the recovery of a 
quantum state $\ket{\Phi}$, undergoing decoherence by the AD channel using the Petz recovery
map. A single-qubit rotation $R_\alpha(\theta)$ along axis $\alpha\in\{x,y,z\}$ with angle $\theta$ is applied to initialize the
system qubit $\rm ^{19}F$ in the 
desired state $\ket{\Phi}$.
Pulse phases are written beneath each rectangle, 
with an overbar indicating
negative phase. The angles $\beta$
and $\delta$ are given by
$\beta=2\cos^{-1}\sqrt{\frac{1+\sqrt{1-p}}{2}}$ and
$\delta=2\cos^{-1}\sqrt{\frac{1+c}{2}}$ with
$c=\sqrt{\frac{1-\epsilon}{1-(1-p)\epsilon}}$. 
The pulse sequence
enclosed within the solid 
box shaded blue, corresponds
to a control-$Z$ operation, while the sequence within the dashed 
box shaded orange, 
represents the control-NOT operations.} 
\label{fig3-ADseq}
\end{figure}

To implement the Petz recovery map with reference state $\sigma=(1-\epsilon)\ket{0}\bra{0}+\epsilon\ket{1}\bra{1}$, we find that the associated unitaries  $U_0',~U_1',$ $\tilde{U}_0'$ and $\tilde{U}_1'$  are  the same as $U_0,~U_1,\tilde{U}_0$ and $\tilde{U}_1$ respectively as in Eq.~\eqref{ad-dqa-unit1}. The unitaries $V'$ and $W'$ appearing in the Petz recovery stage transform as
\begin{equation}
   V'=W'= \begin{pmatrix}
       \sqrt{\frac{1+c}{2}}& \sqrt{\frac{1-c}{2}}\\
       - \sqrt{\frac{1-c}{2}}& \sqrt{\frac{1+c}{2}}
    \end{pmatrix}, \label{ad-dqa-unit2} 
\end{equation}
where $c=\sqrt{\frac{1-\epsilon}{1-(1-p)\epsilon}}$.

\begin{figure}[htb!]
\centering
\includegraphics[width=1\linewidth]{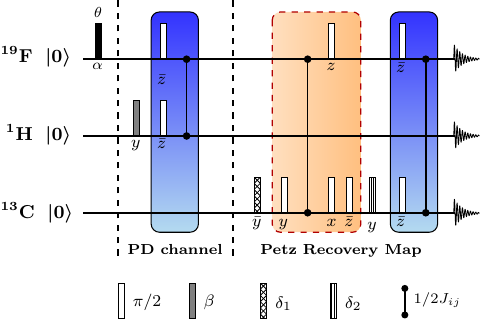}
\caption{(Color online)
NMR pulse sequence to implement a PD channel and the Petz
recovery map on quantum state $\ket{\Phi}$ undergoing 
decoherence under a PD
channel using the Petz recovery map. The system qubit $\rm ^{19}F$ is
initialized in the desired state $\ket{\Phi}$ by applying a single-qubit
rotation $R_\alpha(\theta)$ along the 
axis $\alpha\in\{x,y,z\}$ with an angle
$\theta$. The angles $\beta$, $\delta_1$ and
$\delta_2$ are defined as $\beta=2\sin^{-1}\sqrt{\frac{p}{2}}$ and
$\delta_1=2\cos ^{-1}\sqrt{\lambda_+^2+\mu_+^2}$ and $\delta_2=2\sin
^{-1}\sqrt{\lambda_-^2+\mu_+^2}$. Colored segments indicate two-qubit
operations. The solid rectangular box highlighted in blue
indicates the sequence implementing a control-$Z$ operation, while the
dashed rectangular box shaded in orange denotes the sequence
corresponding to a control-NOT operation.}
\label{fig4-PDseq}
\end{figure}
\subsection {Phase Damping (PD) Channel
and the Corresponding Petz Recovery Map }
\label{sec-pdexp}
Following the same procedure as for the AD channel, we implement the PD channel
using the DQC algorithm. The
Kraus operators are expressed as linear
combinations of Pauli matrices leading to the following unitaries:
\begin{equation}
\begin{split}
     U_0=&~\mathbb{1},\hspace{0.2cm} U_1=Z~, \hspace{0.2cm} \tilde{U}_0=\mathbb{1},\hspace{0.2cm} \tilde{U}_1=\mathbb{1},\\
     V=&\begin{pmatrix}
       \sqrt{1-\frac{p}{2}}&-\sqrt{\frac{p}{2}}\\
       \sqrt{\frac{p}{2}}&\sqrt{1-\frac{p}{2}}
    \end{pmatrix},\hspace{0.2cm} W=\mathbb{1}.        
\end{split}\label{pd-dqa-unit1}
\end{equation} 
 To implement the  Petz map with reference state, $\sigma=(1-\epsilon)\ket{+}\bra{+}+\epsilon\ket{-}\bra{-}$, we obtain the following unitaries
\begin{equation}
    \begin{split}
        U_0'=&~\mathbb{1},\hspace{0.2cm} U_1'=X,\hspace{0.2cm} \tilde{U}_0'=\mathbb{1}, \hspace{0.2cm}\tilde{U}_1'=Z,\\
        V'=&\begin{pmatrix}          
 \sqrt{\lambda_+^2+\mu_+^2} & \sqrt{1-\lambda_+^2-\mu_+^2} \\
 -\sqrt{1-\lambda_+^2-\mu_+^2} & \sqrt{\lambda_+^2+\mu_+^2}
        \end{pmatrix},\\
        W'=&\begin{pmatrix}          
 \sqrt{1-\lambda_-^2-\mu_+^2} & -\sqrt{\lambda_-^2+\mu_+^2} \\
\sqrt{\lambda_-^2+\mu_+^2} & \sqrt{1-\lambda_-^2-\mu_+^2}
        \end{pmatrix}
    \end{split}\label{pd-dqa-unit2}
\end{equation}
where $\lambda_\pm$ and $\mu_\pm$ are defined in Eq. \eqref{pdpetz-kraus}. As
before, $\mathbb{1},X,Z$ represent the identity and Pauli operators,
respectively. For the 
Petz recovery map for PD channel, the decomposition of
the Kraus operator $\mathcal{M}_1^{PD}=Z.(\mu_+\mathbb{1}+\mu_-X)$, involves a
product with the Pauli-$z$ operator. Consequently, a controlled-$Z$ gate is
implemented as controlled $\tilde{U}_1'$.

The NMR pulse sequences for implementing the AD and PD
channels, along with their respective Petz recovery maps,
are shown in Figs.~\ref{fig3-ADseq} and \ref{fig4-PDseq}, respectively. Each
sequence is divided into three segments: the initialization segment, which
prepares the qubit $\rm^{19}F$ in state $\ket{\Phi}$ via a single-qubit
rotation; the damping channel segment, which simulates the decoherence process;
and the recovery segment, which implements the Petz recovery map.

\section{Results and Discussion}

Before discussing the experimental results, we stress that the Petz recovery maps constructed in this work are not expected to provide uniform recovery for arbitrary initial states. Rather, their performance depends sensitively on the overlap between the input state and the reference state used in the construction of the Petz map. This state dependence is an intrinsic feature of Petz recovery and not a
limitation of the experimental implementation. Our experiments explicitly demonstrate this behavior by showing both enhanced recovery and degradation depending on the compatibility between the input state
and the chosen reference state.

To experimentally demonstrate the reference state dependent performance of Petz recovery map  under a damping channel, we performed two sets of experiments for each
input state. In the first set, the system qubit was subjected to the damping
channel (either AD or PD) only. In the second set, following the application of
the damping channel, the associated Petz recovery map was implemented using the
NMR pulse sequence depicted in Figs.~\ref{fig3-ADseq} and \ref{fig4-PDseq}. In
addition, the recovery experiments were repeated for different values of the
reference state parameter $\epsilon$. To evaluate the performance of recovery,
we compared the fidelity of both the damped state and the recovered state with
respect to the original input state of the system qubit, and plotted it as a
function of the strength of the damping channel.

\begin{figure}[htb!]
\centering
\includegraphics[width=1\linewidth]{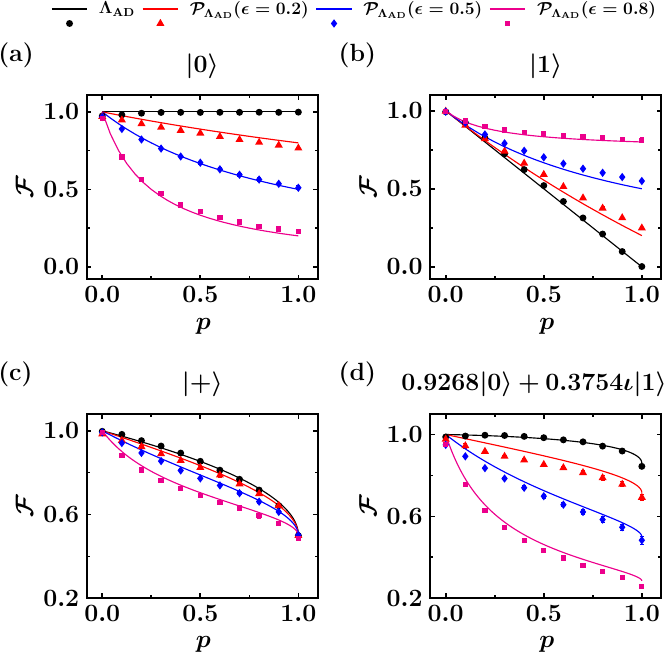}
\caption{(Color online) Experimental results for the AD
channel (black circles) and the corresponding Petz recovery map
implemented using the DQC algorithm. The recovery map is constructed with reference
states of the form
$\sigma=(1-\epsilon)\ket{0}\bra{0}+\epsilon\ket{1}\bra{1}$, for various
$\epsilon$ values. Plots in panels $\rm (a), (b),
(c)$ and $\rm (d)$ depict fidelity $\mathcal{F}$ as a
function of AD channel strength $p$, for initial states
$\ket{0},\ket{1},\ket{+}=\frac{\ket{0}+\ket{1}}{\sqrt{2}}$ and $0.9268
\ket{0} + 0.3754\imath\ket{1}$, respectively. Red triangles, blue
diamonds and magenta squares represent
experimental fidelities after applying the Petz
recovery map with $\epsilon= 0.2, 0.5$ and $0.8$, respectively. The
solid lines depict the corresponding theoretical behavior. The size of
the statistical error bars are  smaller than the  marker size, and are therefore not visible.}
\label{fig5-adplot}
\end{figure}

Fig.~\ref{fig5-adplot} depicts the experimental results for the AD channel and
its corresponding Petz recovery map. For the experimental implementation,
we consider the system qubit initialized in different input states:
$\ket{0},\ket{1},\ket{+}=\frac{\ket{0}+\ket{1}}{\sqrt{2}}$ and $0.9268 \ket{0}
+ 0.3754\imath \ket{1}$. The reference state for constructing the Petz map was
chosen to be of diagonal form,
$\sigma=(1-\epsilon)\ket{0}\bra{0}+\epsilon\ket{1}\bra{1}$, and recovery was
performed for three different values of $\epsilon=0.2,0.5,0.8$. The plots in
the panels $\rm (a)$ and $\rm (b)$, illustrate that under the AD channel,  the
fidelity for the $\ket{0}$ state remains one as expected, since the AD process
drives the system towards the $\ket{0}$ state. Conversely, the
$\ket{1}$ state experiences significant decoherence. On implementing the Petz
recovery map, the fidelity of the recovered state with respect to the $\ket{0}$
state decreases, while  the fidelity with respect to the $\ket{1}$ state
increases,  with increase in $\epsilon$. This behavior
reflects the dependence of the Petz recovery map on the choice of reference
state. For an input state $\ket{0}$, the most suitable reference is one that is
close to $\ket{0}$ itself i.e.  smaller values of $\epsilon$. As $\epsilon$
increases, the reference state shifts toward $\ket{1}$. Thus, for the initial
state $\ket{1}$, higher values of $\epsilon$ are more appropriate. Panels
$\rm(c)$ and $\rm(d)$ display the results for the $\ket{+}$ and $0.9258 \ket{0}
+ 0.3781\iota\ket{1}$ states. Both these states exhibit
behaviors similar to the $\ket{0}$ state. However, the
fidelity loss for the $\ket{+}$ state is less as compared to the $\ket{0}$
state. This occurs because the $\ket{+}$ state is an equal
superposition of the $\ket{0}$ and $\ket{1}$ states.
As a result, the coherence reduces in magnitude, but the state does not
completely collapse to $\ket{0}$. For this state as well, a smaller $\epsilon$
is the better choice.

\begin{figure}
\centering
\includegraphics[width=1\linewidth]{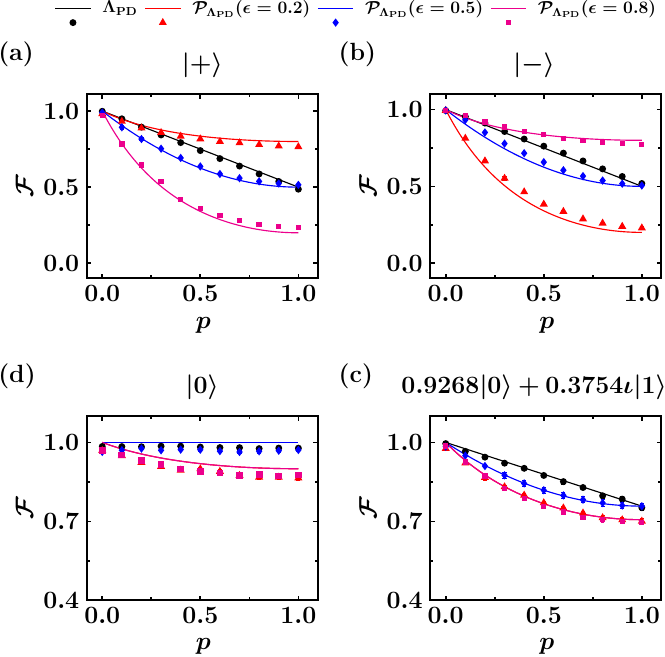}
\caption{(Color online) 
Experimental results showing the effect of the PD
channel and the corresponding Petz recovery map implemented via the DQC algorithm.
The Petz recovery map is designed using
reference states of the form
$\sigma=(1-\epsilon)\ket{+}\bra{+}+\epsilon\ket{-}\bra{-}$, for various
$\epsilon$ values, where $\ket{\pm}$ are eigen state of Pauli-$x$ operator.
Plots in panels $\rm (a), (b), (c)$ and $\rm (d)$ depict fidelity
$\mathcal{F}$ as a function of PD channel strength $p$, for different
initial states: $\ket{+},
\ket{-},\ket{0}$ and $0.9268 \ket{0} +
0.3754\imath\ket{1}$, respectively. Black circles represent fidelity
after phase damping, while red triangles, blue diamonds and magenta
squares show the fidelity after implementing the 
Petz recovery map  with
$\epsilon= 0.2, 0.5$ and $0.8$, respectively. Theoretical predictions
are indicated by solid lines. Statistical error bars are smaller than
the marker size and are hence not visible in the plots.}
\label{fig6-pdplot}
\end{figure}

The experimental fidelity results for system qubit subjected to phase damping channel and its Petz recovery map are depicted in Fig.~\ref{fig6-pdplot}. In this case, we initialize our system qubit in different input states: $\ket{\pm}=\frac{\ket{0}\pm\ket{1}}{\sqrt{2}},\ket{0}$ and $0.9268 \ket{0} + 0.3754\imath \ket{1}$. The Petz map is constructed using reference state  
\(\sigma=(1-\epsilon)\ket{+}\bra{+}+\epsilon\ket{-}\bra{-}\). The experiment is repeated for three different values of $\epsilon=0.2,0.5,0.8$. As highlighted in the previous paragraph, the performance of the Petz map in recovering the damped state strongly depends on the choice of reference state. This behavior under the PD channel is shown in the plots in the panels $\rm (a)$ and $\rm (b)$ of Fig.~\ref{fig6-pdplot}. When $\epsilon$ is small, for e.g. $\epsilon=0.2$, the reference lies closer to state $\ket{+}$ and away from state $\ket{-}$. With this choice of reference state parameter, fidelity of the recovered state increases and we get efficient recovery for the input state $\ket{+}$.  In contrast, the fidelity decreases for the input state $\ket{-}$ (red triangle and red solid line in plots $\rm (a)$ and $\rm (b)$ respectively). Conversely, $\epsilon=0.8$, the reference state moves away from the $\ket{+}$ state and shift towards $\ket{-}$ state. This improves the fidelity for the input state $\ket{-}$, while reducing fidelity for the input state $\ket{+}$  (magenta square and magenta solid line in plots $\rm (a)$ and $\rm (b)$ respectively). Interestingly,  for $\epsilon=0.5$, the reference state becomes maximally mixed, leading to no recovery for both the input states. This is due to the fact that the PD channel specifically affects the off-diagonal elements of the density matrix, and the corresponding Petz map in this case is itself a PD channel. As a result, instead of improving the recovery, the map further reduces the fidelity of the recovered state. For input states $\ket{0}$ and $0.9268 \ket{0} + 0.3754\imath \ket{1}$, the corresponding experimental results are shown in panels $\rm (c)$ and $\rm (d)$, respectively. In case of $\ket{0}$ input state, recovery is not observed for any value of $\epsilon$ except $\epsilon=0.5$. This behavior arises because both the input and the chosen reference state in this case are diagonal in the computational basis, and the PD channel affects only the off-diagonal elements. Although the input state $0.9268 \ket{0} + 0.3754\imath \ket{1}$ is closer in amplitude to the $\ket{0}$ state, the reference states used in constructing the Petz map (based on $\ket{+}$ and $\ket{-}$) are not well suited for recovering this particular input state. As a result, no significant recovery is observed for any of the $\epsilon$ values considered.

The approach employed in this work can, in principle, be extended to larger systems. Within the DQC framework, a quantum channel described by $m$ Kraus operators can be implemented through a coherent linear combination of unitary operations using an ancillary register. In general, the number of ancillary qubits required scales as $\lceil \log_2 m \rceil$, allowing the implementation of channels with increasingly complex Kraus structures. Consequently, the same strategy used here for single-qubit damping channels can be generalized to multi-qubit noise processes and higher-dimensional systems. The primary experimental challenge lies in the implementation of the required controlled unitary operations and the corresponding increase in circuit depth. Nevertheless, the framework provides a systematic route for studying recovery maps and noise-reversal protocols in larger quantum systems.

\color{black}
\section{Conclusions} 
\label{concl}
We have experimentally investigated the performance of the Petz recovery map in
retrieving the quantum state of a single qubit undergoing  amplitude damping
and phase damping on an NMR quantum information processor. By initializing the
qubit in different input states and implementing both channels and their
associated Petz recovery maps via the duality quantum computing algorithm, we analyzed how
recovery fidelity depends on the choice of the reference state used in the
construction of the Petz map. Our results show that the performance of recovery
is highly sensitive to the overlap between the input state and the reference
state. For the AD channel, we employed a diagonal reference state.  Since the
AD channel naturally drives any input state towards the $\ket{0}$ state, the
recovery map was found to be most effective for the $\ket{1}$ state for all
choices of the input state that we considered. In contrast, for the PD channel,
we used a reference state that is in a parameterized superposition of $\ket{+}$
and $\ket{-}$ states. In this case, the recovery fidelity for the input state
(for e.g. $\ket{+}$) improves with increasing overlap with the reference state,
while it decreases for the orthogonal one (for e.g.  $\ket{-}$), under the same
conditions.

Beyond the specific implementation presented here, our results provide an experimental platform for investigating the recoverability of quantum information under realistic noise processes. Since the Petz map represents a canonical recovery operation associated with a given noise channel, its implementation allows one to experimentally probe the limits of noise reversibility in controlled quantum systems. More broadly, the approach illustrated here can be extended to other quantum channels and larger systems, offering a systematic way to explore channel-adapted recovery strategies and their relation to noise mitigation and quantum error-correction concepts. Such studies may help clarify the fundamental limits of information recovery in open quantum systems and provide useful insights for the development of channel-adapted noise mitigation strategies in near-term quantum technologies.
\color{black}
\acknowledgements
All experiments were performed on a Bruker Avance-III 600 MHz FT-NMR
spectrometer at the NMR Research Facility at IISER Mohali. G.S. acknowledges University Grants Commission (UGC), India, for financial support. N.K.B. acknowledges financial support from CNPq Brazil (442429/2023-1) and FAPESP (Grant 2021/060350). N.K.B. is part of the Brazilian National Institute for Quantum Information (INCT Grant 465469/2014-0).

\end{document}